\documentclass[12pt]{article}

\usepackage{epsfig,latexsym,amsfonts,amsmath,amsthm,amssymb,amsbsy,multirow,slashed,wasysym,textcomp,dsfont,comment,mathtools,cancel,cite,diagbox,datetime,appendix,BOONDOX-calo}
\usepackage{tocloft}
\usepackage{bbold}
\usepackage{sidecap}
\usepackage{adjustbox}
\usepackage{pgfplots}
\usepackage{pgfkeys}
\usepackage{oplotsymbl}
\usepackage{tikzpagenodes}
\usepackage{adforn}
\tikzstyle{bag} = [align=center]
\usetikzlibrary{shapes,backgrounds,arrows.meta,patterns,decorations.markings,bending,positioning, decorations.pathmorphing,cd} 
\tikzset{snake it/.style={decorate, decoration=snake}}
\usepackage{adjustbox}
\usepackage[normalem]{ulem}
\usepackage[mathscr]{eucal}
\usepackage[version=3]{mhchem}
\usepackage{physics}

\usepackage[hidelinks]{hyperref}
\usepackage{subcaption}

 \newcommand{\badat}{\begin{alignedat}}
 \newcommand{\eadat}{\end{alignedat}}
 \newcommand\scalemath[2]{\scalebox{#1}{\mbox{\ensuremath{\displaystyle #2}}}}
 \def\be{\begin{equation}}
\def\ee{\end{equation}}

\def\p{\partial}

\usepackage{color}

\newcommand{\pink}[1]{\textcolor{\pink}{#1}}

\definecolor{dblue}{rgb}{0.2,0.50,0.80}

\usepackage[utf8]{inputenc}

\tikzset{snake it/.style={decorate, decoration=snake}}

\def\bz{{\bar z}}

\usepackage{tensor}
\usepackage{amsmath}
\usepackage{amssymb}
\usepackage{mathrsfs}
\usepackage[normalem]{ulem}
\usepackage{bbm} 
\usepackage{graphicx}
\usepackage{stmaryrd}
\usepackage{xcolor}
\usepackage{mathtools}
\usepackage{xfrac}

\renewcommand{\bz}{\bar z}

\def\bz{{\bar z}}

\def\scri{\mathscr I}

\numberwithin{equation}{section} 
\pgfplotsset{compat=1.17} 
\begin{document}

\begin{titlepage}
\thispagestyle{empty}
  \begin{flushright}
  \end{flushright}
  \bigskip

  \begin{center}

                  \baselineskip=13pt {\LARGE \scshape{
                A Comment on Boundary Correlators: \\ [.5em]
                 Soft Omissions and the Massless S-Matrix
            }
         }

      \vskip1cm

   \centerline{  Eivind J\o{}rstad$^{1,2}$ and {Sabrina Pasterski}$^1$ }

\bigskip\bigskip
 \bigskip

 \centerline{\em${}^1$  
 Perimeter Institute for Theoretical Physics, }
 \vspace{0.2em}
 \centerline{\em Waterloo, ON N2L 2Y5, Canada}

\vspace{1em}

\centerline{\em${}^2$  
Dept. of Physics \& Astronomy, University of Waterloo,}
\vspace{0.2em}
\centerline{\em Waterloo, ON N2L 3G1, Canada}

\bigskip\bigskip

\end{center}

\begin{abstract}
We revisit the extrapolate dictionary for massless scattering in flat spacetime and identify a soft contribution that is typically dropped from the saddle point approximation. We show how to consistently regulate the extrapolation to include both the soft and hard components and identify the boundary correlation functions as a combination of electric and magnetic branch Carrollian correlators. This implies in particular that there are contributions to these boundary correlators that are non-distributional on the celestial sphere. Finally, we close by exploring the utility of the magnetic branch for extracting celestial data from low point correlators: connecting our results to recent work on flat space extrapolate dictionaries and celestial shadow amplitudes.

\end{abstract}

 \bigskip \bigskip \bigskip \bigskip

\end{titlepage}

\setcounter{tocdepth}{2}

\tableofcontents

\section{Introduction}

Generalizing the AdS/CFT correspondence to asymptotically flat spacetimes is an adventure riddled with subtleties. The celestial holography program encompasses a recent effort to tackle this question, making headway by exploiting symmetry enhancements that appear for such geometries~\cite{Pasterski:2021raf,Strominger:2017zoo,Pasterski:2021rjz,Pasterski:2023ikd,Raclariu:2021zjz,McLoughlin:2022ljp,Donnay:2023mrd}. The main proposal is that quantum gravity in asymptotically flat spacetimes can be encoded in a conformal field theory living on the codimension 2 celestial sphere. The central objects of study are celestial amplitudes which are S-matrix elements prepared in a basis of boost eigenstates. For massless particles, this change of basis is implemented by a Mellin transform in the particle's energy~\cite{deBoer:2003vf,Pasterski:2017ylz}.
By another change of basis the conformal dimension can be traded for a null time coordinate $u$ and the massless sector of the celestial CFT can be recast as a Carrollian CFT living on null infinity~\cite{Donnay:2022sdg,Donnay:2022wvx}.

The proposed boundary CFT is somewhat exotic at first glance. In the celestial description, the weights that capture a basis of single particle states span the principal series and hence are unlike the usual unitary 2D CFTs we are accustomed to~\cite{Pasterski:2016qvg,Pasterski:2017kqt,deBoer:2003vf}. Meanwhile the Carrollian description also has exotic looking correlators~\cite{Donnay:2022aba,Bagchi:2022emh, Donnay:2022wvx,Saha:2023hsl,Bagchi:2023cen}, despite having the usual codimension. Namely, for the same reason that translation invariance in the bulk force the momentum space amplitude to contain a momentum conserving delta function, the low point celestial correlators we are interested in are distributional on the celestial sphere~\cite{Pasterski:2017ylz,Mizera:2022sln}. These singular correlation functions are not what one expects of a regular CFT, and are not what one expects from the naive flat limit of AdS/CFT.\footnote{See \cite{Alday:2024yyj} for a recent study of how to extract electric and magnetic branch Carrollian correlators from the flat limit of AdS CFT that appeared while this work was in preparation. There only one or the other will appear. For perturbative computations in the free theory, we are interested in the edge case where $\Delta=1$ in their (4.5).} This has lead to various attempts to try to change the scattering basis to smear out these singularities of the CCFT~\cite{Crawley:2021ivb,Fan:2022vbz,Sharma:2021gcz,Banerjee:2022hgc,Hu:2022syq,Jorge-Diaz:2022dmy,Chang:2022jut,Furugori:2023hgv}, as well as recent efforts to look at analytic terms that appear as contributions from individual leaves of a hyperbolic foliation of the bulk spacetime~\cite{Melton:2023bjw,Melton:2024jyq,deBoer:2003vf,Cheung:2016iub}.

In this paper we revisit the extrapolate dictionary for massless particles~\cite{Pasterski:2021dqe,Donnay:2022sdg,He:2020ifr,He:2014laa,Strominger:2017zoo,Jorstad:2023ajr} and show that there is a zero energy soft term that is usually omitted from the saddle point approximation. We show that dropping this zero-mode is consistent with doing LSZ to get the amplitude. In other words amplitude $\neq$ boundary correlator.  This term is non-distributional but would seemingly be less interesting to people wanting to study scattering precisely because it is a strictly soft mode. However, we will find that for the low point correlators it gives us a bit of mileage. It contains information about the dynamics. Furthermore, it will let us make contact between certain analytic terms that have appeared in recent studies of flat space dictionaries~\cite{Kim:2023qbl,Jain:2023fxc,Banerjee:2024yir} and the celestial and Carrollian constructions. 
It clarifies that celestial correlators and celestial amplitudes need not be one and the same. In particular, certain choices have to be made when picking which `branch'~\cite{Mason:2023mti,Salzer:2023jqv,Bagchi:2023fbj,Donnay:2022wvx} of Carrollian correlators are relevant to flat holography. Up to now these have mainly been designed to match the amplitudes rather than the full extrapolated correlator.

This paper is organized as follows. In section~\ref{sec:softadditions} we identify a soft term that is dropped in the usual saddle point approximation. We then turn to the two point functions in section~\ref{sec:carrollbranches}, computing the corrected extrapolate two-point functions in~\ref{sec:two_point} and reinterpreting this corrected dictionary as giving a combination of electric and magnetic branch Carrollian correlators in section \ref{sec:em_branch}. As we explore in section~\ref{sec:dynamics}, this result lets us connect the distributional celestial amplitudes with the analytic correlators appearing in~\cite{Jain:2023fxc}. We first review how the soft terms do not change the finite energy amplitudes in section \ref{sec:LSZ}, before exploring how the dynamics is encoded in the low point magnetic branch correlators in section \ref{sec:soft4pt}. We then connect our results to recent work on shadow celestial amplitudes in section~\ref{sec:shadow}.  
We leave some useful but tangential computations to the appendix: filling out some details of how the 2-point correlator is reproduced by the two terms in the corrected extrapolate dictionary in section~\ref{app:extrapolatcheck}, and then setting up the extrapolate dictionary with different choices of regulating cutoff surfaces in appendix \ref{app:extrapolate}.  This facilitates comparing results that appear in recent papers on various flat holography dictionaries~\cite{He:2020ifr,Kim:2023qbl,Jain:2023fxc,Kraus:2024gso}.

\section{Restoring Soft Terms in the Extrapolate Dictionary}\label{sec:softadditions}

As mentioned above, many references define CCFT correlation functions in terms of bulk scattering amplitudes, up to a change of basis see e.g~\cite{Pasterski:2016qvg,Pasterski:2017ylz,He:2014laa,Strominger:2017zoo}. This leads to these correlation functions having unfamiliar properties as compared to a standard CFT. For example, such correlators must be proportional to a momentum conserving delta-function. Below, we explain why the identification of the amplitude with the correlation function omits a soft term. In this section, we will correct this mistake in defining the extrapolated operators, before examining their correlation functions and comparing them to Carrollian correlators in section~\ref{sec:carrollbranches}. 

For simplicity, consider a free massless scalar field $\phi$ in $d=4$ dimensional Minkowski space. We will add interactions in section~\ref{sec:dynamics}. We want to push a field operator to future null infinity along some fixed null hypersurface.
It will be useful to work in flat-Bondi coordinates~\cite{He:2020ifr}
\be
x^\mu=\frac{r}{2}\left(1+|z|^2+\frac{u}{r},z+\bz,-i(z-\bz),1-|z|^2-\frac{u}{r}\right)
\ee
for which the metric takes the form
\begin{equation}\label{eq:flat-bondi}
    ds^2=-du\,dr+r^2dz\,d\bar{z},
\end{equation}
where $-\infty<r<\infty$, with the boundaries corresponding to past and future null infinity. Furthermore, we will
parameterize null momenta by
\begin{equation}
\begin{split}
    p^{\mu}=\omega&\left(1+|z|^2,z+\bar{z},-i(z-\bar{z}),1-|z|^2 \right).
    \label{eq:momentumpar}
\end{split}
\end{equation}
 We would like to consider the limit of sending our massless field operators to null infinity 
\be\begin{split}
\Phi(u,z,\bar{z})\equiv\lim_{r\to\infty}r^{\delta}\phi (x^\mu)|_{u=\text{const}}&=\lim_{r\to\infty}\left[r^{\delta}\int\limits \frac{\text{d}^{3} p}{(2\pi)^{3} 2\omega_p }a_p e^{ i p\cdot x}+\text{h.c.}\right]\bigg|_{u=\text{const}}
\end{split}
\label{eq:begin}
\ee 
where $u$ is a null coordinate and we have scaled the operator by the appropriate power of $r$ to obtain a finite boundary result. Note, for the free massless scalar in four spacetime dimensions $\delta=1$. The limit on right hand side of \eqref{eq:begin} is commonly extracted from the saddle point approximation~\cite{He:2014laa} to give the operators appearing in celestial amplitude $\langle\mathcal{O}_1...\mathcal{O}_n\rangle$, where $\mathcal{O}$ is given by
\be
\mathcal{O}(u,z,\bz)\equiv-\frac{i}{4 \pi^2 }\int\limits_0^{\infty}d\omega\, a(\omega,z,\bar{z})\, e^{- i \omega u 
}+\text{h.c.}.
\label{eq:approx}
\ee
From the saddle point approximation introduced in~\cite{He:2014laa,Strominger:2017zoo}, one would expect \eqref{eq:begin} and \eqref{eq:approx} to be proportional to one another, so that we just need to Fourier-transform in the null coordinate $u$ to obtain the usual creation or annihilation operator. As such one is tempted to identify correlation functions of $\Phi$'s with the S-matrix. In other words, one naively expects the $\mathcal{O}$ and $\Phi$ correlators to be the same.
 
However, below we will compare the correlation functions of such extrapolated bulk operators with the extrapolated bulk correlation functions given schematically by
\begin{equation}
    \lim_{r\to\infty} r^{n\delta} \langle\phi(x_1)...\phi(x_n)\rangle.
    \label{eq:begin2}
\end{equation}
We will show that the usual procedure used to evaluate~\eqref{eq:approx} gives answers inconsistent with~\eqref{eq:begin2}. Let us begin by identifying where the original calculation goes wrong and how to fix it.

\paragraph{Taking the limit}

Before taking the limit \eqref{eq:begin} in flat-Bondi coordinates $(u,r,z,\bar{z})$ let us change our  
\begin{equation}
\openup 3\jot
\begin{split}
         r\phi(u,r,z,\bar{z})&=\frac{ r}{(2\pi)^3}\int\limits_0^{2\pi}d\theta\int\limits_0^{\infty} d\rho \int\limits_0^{\infty}d\omega \, \left(\omega\,a_p e^{-i \omega u}\right)e^{-i r\omega \rho} +\text{h.c.},\\
         \vspace{1cm}
        &\text{where}\quad\rho=|z-z'|^2,\qquad \theta=\text{arg}(z-z').
\end{split}
\end{equation}
Note that, except for the prefactor, $r$ only appears in an exponential that is symmetric in $\rho$ and $\omega$, hinting that the usual argument should give an answer that is also in some sense symmetric in those variables. To make this more explicit, let us again change coordinates to 
\begin{equation}
    \frac{\omega}{\omega_0}= \frac{\sqrt{ j+s^2}+s}{2},\qquad \rho=\frac{\sqrt{ j+s^2}-s}{2},\quad j\geq 0,\,\,-\infty<s<\infty
\end{equation}
such that $\frac{\omega}{\omega_0} \rho = j/4 $ and $\omega/\omega_0-\rho=s$. For now $\omega_0$ is an arbitrary constant introduced for dimensional reasons. However, below we will see that it makes sense to identify $\omega_0$ as a UV scale. To avoid clutter, let us also introduce the notation
\begin{equation}
    F(\omega,\rho)\equiv \int\limits_0^{2\pi}d\theta \,\omega\,a_p e^{-i \omega u}.\label{eq:Ffunc}
\end{equation}
In these new coordinates we thus get 
\begin{equation}\scalemath{.95}{
      r\phi(u,r,z,\bar{z})=\frac{r \omega_0}{4(2\pi)^3}\int\limits_0^{\infty} dj \int\limits_{-\infty}^{\infty}d s \, \frac{1}{\sqrt{ j+s^2}}F(\omega_0 \frac{\sqrt{ j+s^2}+s}{2},\frac{\sqrt{ j+s^2}-s}{2})e^{-i r \omega_0 j/4  } +\text{h.c.},}
      \label{eq:betterex}
\end{equation}
which is an expression for which we can straightforwardly take the  large $r$ limit.  By integration by parts in $j$ we see that the contribution leading in $r$ comes from $j=0$. Meanwhile the $j\rightarrow\infty$ contribution can be discarded. This is the same argument as that used to derive \eqref{eq:approx}, but now applied to the more appropriate variable $j$. We thus find
\begin{equation}
\begin{split}
      (2\pi)^3\,\Phi(u,z,\bar{z})&= \int\limits_{-\infty}^{\infty}d s \, \frac{1}{i|s|}F\left(\omega_0 \Theta(s) s,-\Theta(-s)s\right) +\text{h.c.}\\
        &=\int\limits_{0}^{\infty}d s \, \frac{F( \omega_0 s,0)}{i s}-\int\limits_{0}^{\infty}d s \, \frac{F( 0,s)}{i s} +\text{h.c.}\\
        &=-  2\pi i\int\limits_0^{\infty}d\omega\, a(\omega,z,\bar{z})\, e^{- i \omega u}+ i \int d^2 z' \,\frac{\omega a(\omega,z',\bar{z}')|_{\omega=0}}{|z-z'|^2}+\text{h.c.},
        \label{eq:corrected expression}
\end{split}
\end{equation}
which equals the original approximation \eqref{eq:approx} plus an additional term which smears over the zero mode.\footnote{We will discuss it's interpretation as a shadow transform in section~\ref{sec:dynamics}.} Note that $\omega a(\omega,z,\bar{z})|_{\omega=0}$ does not vanish when there is an infrared pole. We also note that our coordinate system is such that this expression is the same as the one for an incoming particles at the antipodal point. 

Both integrals on the right hand side of~\eqref{eq:corrected expression} can yield divergences. For instance, as we discuss below, dropping the soft terms would lead to the conclusions that the $\langle \Phi_1 \Phi_2 \rangle$ has an infrared divergence. We observe that the singular behaviour in one term can be cancelled against the other. One way to see this is by integrating by parts using $s^{-1}=\frac{d}{ds}(\log s)$. We can throw away the contributions from $s\rightarrow\infty$ by  assuming a $u$ is appropriately analytically continued to suppress the UV part of the integral. Let us also assume we can throw away the $s\rightarrow-\infty$, i.e.~large $\rho$ contribution, which leaves us with
\begin{equation}
\begin{split}
    (2\pi)^3\,\Phi(u,z,\bar{z})=&\int\limits_{0}^{\infty}d s \, \log(s)\,i \frac{d}{ds}F( \omega_0 s,0)-\int\limits_{0}^{\infty}d s \, \log(s)\, i \frac{d}{ds}F( 0,s) +\text{h.c.}\\
    =&\, 2\pi i \int\limits_{0}^{\infty}d \omega \, \log(\frac{\omega}{\omega_0})\, e^{-i \omega u}\left(\partial_\omega (\omega a(\omega,z,\bar{z}))-i \omega u \,a(\omega,z,\bar{z})\right) \\
    &-i\int d^2 \tilde z \, |\tilde z|^{-1} \log|\tilde z|\,  \left(\omega \partial_{|\tilde z|}a(\omega,\tilde z+z,\bar{\tilde z}+\bar{z})\right)\big|_{\omega=0} +\text{h.c.},
    \label{eq:extrapop}
\end{split}
\end{equation}
where $\tilde z \equiv z'-z$. We note that the logarithm has an integrable singularity at zero. The above expression makes it explicit that the extrapolate operator does not give rise to infrared divergences even in the presence of an infrared pole.

\section{Identifying `Mixed-Branch' Carrollian Correlators}\label{sec:carrollbranches}

Now that we have determined the corrected extrapolate operators we can now examine their correlation functions. In this section we will focus on the free theory, evaluating the two point functions for the extrapolated operators in section~\ref{sec:two_point}, before identifying these as a combination of the electric and magnetic branch Carrollian correlators in section~\ref{sec:em_branch}.

\subsection{Two-Point Functions of Corrected Extrapolate Operators}\label{sec:two_point}

Consider the free massless scalar in 4D. There are two ways one might naively try to evaluate the extrapolate dictionary correlators. First, lets start with the bulk to bulk Feynman propagator in position space. This is just the time ordered two point function, which takes the following form
\begin{equation}\label{eq:pm}
    \langle T\!\left\{ \phi(x_1)\phi(y_1) \right\} \rangle  = \frac{1}{4\pi^2}\frac{1}{(x_1-x_2)^2+ i \epsilon}\,
\end{equation}
where the $i\epsilon$ gives the appropriate time ordering. Let us now evaluate this correlator on a fixed $r$ hypersurface and take $r\rightarrow \infty$. 
We will stick to the null hyperplane case with one $in$ and one $out$ operator, which will make the time ordering trivial.\footnote{See appendix~\ref{app:extrapolate} for a discussion about other regulators and subtleties about $in$-$in$ and $out$-$out$ contact terms for the case of a null cutoff surface. The latter are interesting for studying vacuum wavefunctionals as in~\cite{Jain:2023fxc}.} 
Taking the naive large $r$ limit gives
\begin{equation}\label{eq:noncontact}
    \lim_{r\to\infty}  \langle T\{r_1\phi(x_1) r_2\phi(x_2)\} \rangle|_{r_1=-r_2=r,z_1\neq z_2} = \frac{1}{4\pi^2}\frac{1}{|z_1-z_2|^2}
\end{equation}
which has no $u$-dependence. This is particularly bad if we want to Fourier transform to momentum space, since it would seem to imply vanishing two point functions at non zero energy. So far we have assumed $z_1\neq z_2$. Let's see how to extract any contact terms.

The route that would immediately land on the contact term contribution in the celestial literature is as follows. Starting from the single particle states in momentum space
\be
|p\rangle =a^\dagger_p |0\rangle
\ee
the canonical commutation relations
\begin{equation}
    \left[a(\omega_1,z_1,\bar{z}_1),a^{\dagger}(\omega_2,z_2,\bar{z}_2)\right]=(2\pi)^3\,\frac{1}{2 \omega_1} \delta(\omega_1-\omega_2)\,\delta^{2}(z_1-z_2)
\end{equation}
imply the following inner product
\be
\langle p_1|p_2\rangle= (2\pi)^3\,\frac{1}{2 \omega_1} \delta(\omega_1-\omega_2)\,\delta^{2}(z_1-z_2).
\ee
To evaluate the Carrollian correlators, one can try to Fourier transform back to position space, defining
\begin{equation}
    \langle \mathcal{O}(u_1,z_1,\bar{z_1})\mathcal{O}(u_2,z_2,\bar{z_2})\rangle = -\frac{\delta^{2}(z_1-z_2)}{4\pi}\int\limits^\infty_0 d\, \omega  \frac{e^{-i\omega(u_1-u_2)}}{\omega}.
\end{equation}
This is a somewhat pathological integral, but it can be dealt with, e.g.~following \cite{Donnay:2022wvx}. In particular, we can introduce an infrared cutoff to find 
\begin{equation}
    \tilde{C}(u)\equiv\int\limits^\infty_{\omega_{IR}} d\, \omega  \frac{e^{-i\omega u}}{\omega}=-\gamma-\log\left(\omega_{IR}\, |u|\right) - \frac{i \pi}{2}\text{sign}\left(u\right)+O(\omega_{IR}),
    \label{eq:IRdivversion}
\end{equation}
where $\gamma$ is the Euler-Mascheroni constant. 
We thus find
\begin{equation}
    \langle \mathcal{O}(u_1,z_1,\bar{z_1})\mathcal{O}(u_2,z_2,\bar{z_2})\rangle = -\frac{1}{4\pi} \tilde{C}(u_1-u_2) \delta^{(2)}(z_1-z_2),
\end{equation}
for the boundary 2-point function. In other words, this two-point function is an angular delta function with an IR divergent coefficient.\footnote{Again, because one operator is at $\mathcal{I}^-$ and the other is at $\mathcal{I}^+$ this is already the time ordered correlator with respect to the usual bulk time coordinate, rather than the null $u$-coordinate. This correlator has support only at antipodal points, and we can understand the step function in $u$ as signaling whether the two operators are timelike or spacelike separated. 
} Thus, the usual celestial 2-point function does not capture the non-distributional part of the bulk correlation function seen in \eqref{eq:noncontact}.

Let us now keep track of what happens at $z_1=z_2$. 
The full position space expression is
\begin{equation}\label{eq:+-full}
     \langle T\{r_1\phi(x_1) r_2\phi(x_2)\} \rangle|_{r_1=-r_2=r}=\frac{1}{4\pi^2}\frac{1}{|z_1-z_2|^2+\frac{2(u_1-u_2)}{r}+i \epsilon r^{-2}}.
\end{equation}
As long as $z_1\neq z_2$ the limit will be identical to~\eqref{eq:noncontact}. However, if $z_1=z_2$, the expression blows up in a $u$-dependent way. Correspondingly, the expression~\eqref{eq:+-full} has a $u$-dependent contact term at $z_1=z_2$
\begin{equation}
    \lim_{r\to\infty}  \langle T\{r_1\phi(x_1) r_2\phi(x_2)\} \rangle|_{r_1=-r_2=r} = \frac{1}{4\pi^2}\frac{1}{|z_1-z_2|^2} +\frac{1}{4\pi}  C(u_1-u_2)\,\delta^{(2)}(z_1-z_2)
    \label{eq:fullextrapprop}
\end{equation}
where we can obtain the coefficient of the delta function by taking the difference of \eqref{eq:+-full} and the first term of \eqref{eq:fullextrapprop}\footnote{Note, one has to keep the dimensionless regulator $\epsilon r^{-2}$ when taking the integral to obtain a finite expression.} and integrating over $z$ to find
\begin{equation}\label{eq:UVdivversion}
    C(u)= -{\log(\frac{2 r}{\epsilon}|u|)-i \frac{\pi}{2}} \text{sign} (u)+O(r^{-1}).
\end{equation}
Thus, we get a combination of contact and power law terms which we will identify as a combination of
electric and magnetic branch correlators in the next subsection. Note the difference between the contact term coefficients in the usual celestial 2-point function \eqref{eq:IRdivversion} and the position space expression \eqref{eq:UVdivversion}: while they only differ by an additive constant, the former the constant diverges to positive infinity for $\omega_{IR}\to 0$, while the latter diverges to negative infinity for $r/\epsilon\to \infty$.

Satisfyingly, as is shown in Appendix \ref{app:extrapolatcheck}, we can reproduce both the contact and analytical part if we evaluate vacuum correlators of our corrected extrapolate dictionary operators from section~\ref{sec:softadditions}. We can also choose the constant $\omega_0$ in terms of $r$ and $\epsilon$ (namely, $\omega_0\sim r/\epsilon$ -- see the appendix) to reproduce the right divergent constant term, suggesting that $\omega_0$ should be thought of as a UV frequency scale. In other words, we are able to restore the expected relationship 
\begin{equation}
\langle \Phi(u_1,z_1,\bz_1)\Phi(u_2,z_2,\bz_2) \rangle =   \lim_{r\to\infty}  \langle T\{r_1\phi(x_1) r_2\phi(x_2)\} \rangle|_{r_1=-r_2=r},
\end{equation}
which one would originally have expected to be true for an extrapolated operator.

\subsection{Electric and Magnetic Branch Carrollian Correlators}\label{sec:em_branch}

Let's now give some context for these correlators within the literature on Carrollian CFTs, in particular as applied to flat holography~\cite{Bagchi:2016bcd,Ciambelli:2018wre,Donnay:2022aba,Bagchi:2022emh, Donnay:2022wvx,Saha:2023hsl,Bagchi:2023cen}. Such CFTs live on a null manifold whose global symmetry group is Poincar\'e.  Now one subtlety is that there are multiple solutions to the Ward identities enforcing Poincar\'e covariance of the correlators. These have gone by different names in the literature, ex. time independent vs dependent branches in~\cite{Donnay:2022wvx} and CFT versus delta function branches in~\cite{Bagchi:2023fbj}. The reasons for these namings will be obvious from their form in what follows. Here we will refer to the two solutions as electric and magnetic branches~\cite{deBoer:2023fnj}, matching the conventions in the recent works \cite{Salzer:2023jqv,Mason:2023mti,Alday:2024yyj}. Different actions that give rise to such correlators on the boundary have been studied in~\cite{Campoleoni:2022ebj,Ciambelli:2023xqk,deBoer:2023fnj}. Here we are more interested in how they arise from the bulk.

A Carrollian primary of weight $(k,\bar k)$ transforms as follows~\cite{Donnay:2022aba}
\be\label{carprim}
    \delta_\xi\Phi_{(k,\bar k)}(u,z,\bz)
        := [(f+u\alpha)\p_u+Y\p+k\p Y +\bar Y{\bar\p}+\bar k\bar \p \bar Y ]\Phi_{(k,\bar k)}(u,z,\bz)\,,
\ee
under global translations 
\be\label{eq:tr}
    f_0=1+z\bz\,,\quad 
    f_1=-(z+\bz)\,,\quad
    f_2=-i(\bz-z)\,,\quad
    f_3=-1+z\bz\,,
\ee
and Lorentz transformations 
\be\begin{array}{lll}\label{lorentz}
    Y^z_{12}=iz\,,\quad
        & Y^z_{13}=-\frac{1}{2}(1+z^2)\,,\quad
        & Y^z_{23}=-\frac{i}{2}(1-z^2)\,,\\
   Y^z_{03}=z\,,\quad
        & Y^z_{02}=-\frac{i}{2}(1+z^2)\,,\quad
        & Y^z_{01}=-\frac{1}{2}(1-z^2)\,.
\end{array}
\ee
As in usual CFTs, the form of the low point functions is nearly fully determined by demanding covariance under the global conformal group. However, one subtlety is that the solutions to the Poincar\'e Ward identities include distributional ones which are relevant to scattering amplitudes, while the ones that are non-distributional on the celestial sphere have trivial time dependence. We will turn to their low point functions, in turn.

\paragraph{Electric Branch} The electric branch, or time-dependent, correlators have the following two point functions~\cite{Donnay:2022wvx}
\begin{equation}\label{ebranch}
\left\langle\Phi_{\left(k_1, \bar{k}_1\right)}\left(x_1\right) \Phi_{\left(k_2, \bar{k}_2\right)}\left(x_2\right)\right\rangle=\frac{c_E}{\left(u_1-u_2\right)^{k_{12}^{+}-2}} \delta^{(2)}\left(z_1-z_2\right) \delta_{k_{12}^{-}, 0}
\end{equation}
where
\be
k_{12}^{\pm}=a\pm b,~~a=\sum_i k_i ,~~ b=\sum_i \bar{k}_i.
\ee
Note that for a scalar field in the bulk we expect $k_i=\bar k_i=\frac{1}{2}$ and~\eqref{ebranch} will develop a logarithmic $u$ dependence as $k_{12}^+\rightarrow 2$. Indeed when $k_{12}^+=2+n$ for $n\in\mathbb{N}$ there are additional distributional solutions that satisfy the Carrollian Ward identities that involve various derivatives of ${\rm sign}(u_{12})$~\cite{Donnay:2022wvx}.

\paragraph{Magnetic Branch}

Meanwhile magnetic branch correlators look more like conventional CFT correlators on the celestial sphere,  while being time independent~\cite{Donnay:2022wvx}
\begin{equation}
\left\langle\Phi_{\left(k_1, \bar{k}_1\right)}\left(x_1\right) \Phi_{\left(k_2, \bar{k}_2\right)}\left(x_2\right)\right\rangle=\frac{c_M}{\left(z_1-z_2\right)^{k_1+k_2}\left(\bar{z}_1-\bar{z}_2\right)^{\bar{k}_1+\bar{k}_2}} \delta_{k_1, k_2} \delta_{\bar{k}_1, \bar{k}_2}.
\end{equation}
This $u$-independence means that they only appear in the infrared limit of boundary correlators. For this reason the electric rather than magnetic branch is what have been matched to scattering amplitudes in~\cite{Donnay:2022aba,Donnay:2022wvx}. Meanwhile, both branches have been examined within the flat limit of AdS/CFT in~\cite{Alday:2024yyj}.  

\vspace{1em}
\noindent Comparing to our results above, we see that the true extrapolate dictionary should involve a linear combination of electric and magnetic branches. In~\cite{Alday:2024yyj}, the different branches are seen to appear with different powers of the AdS scale that depend on the dimension of the operator. Upon further inspection, one can see that for the case of a scalar field in the bulk both terms should contribute, consistent with our corrected extrapolate dictionary.

 \section{Implications of Restoring the Soft Terms}\label{sec:dynamics}

We will conclude by considering the ramifications of restoring the previously omitted soft terms in our extrapolate dictionary. First, we will show how the procedure for extracting the S-matrix is not modified in section~\ref{sec:LSZ}, but that these magnetic branch correlators still capture dynamics in section~\ref{sec:soft4pt}. Finally, we will close by connecting this story to the literature on shadow celestial amplitudes in section~\ref{sec:shadow}.

\subsection{Soft Modes and the S-matrix}\label{sec:LSZ}

In this section, we will consider what happens when we turn on perturbative interactions. We propose that applying the LSZ formula has the effect of throwing away the soft contributions we found above. Thus, if one had neglected the soft terms from the beginning, correlation functions of the extrapolated operator would directly compute the amplitude.

 We will use $\lambda \phi^4$ theory as a concrete example, which will also be convenient for comparing our discussion to recent results in~\cite{Jain:2023fxc} and~\cite{Alday:2024yyj}.
More general discussions of LSZ and boundary correlation functions can be found in recent papers such as~\cite{He:2020ifr,Pasterski:2021dqe, Kim:2023qbl,Jain:2023fxc,Kraus:2024gso}. We start with the familiar form of the LSZ formula
\begin{equation}
    \mathcal{A}(\{x_i,y_j\})=\left(\prod_{i\in\textit{in}}\int d^4 x_i \, e^{-i p_i\cdot x_i }\nabla_i ^2\right)\left(\prod_{j\in\textit{out}}\int d^4 y_j \, e^{i p_j\cdot y_j }\nabla_j ^2\right)\langle T \{\phi(x_i), \phi(y_j)\}\rangle,
\end{equation}
where the indexes $i$ and $j$ label incoming and outgoing particles respectively. As in \cite{Jain:2023fxc}, we will consider a cut off spacetime, though we here choose a large $\pm r$, null cutoff surface in flat Bondi coordinates, rather than a cutoff in early and late Lorentzian time.

 Having chosen a cutoff surface, we can integrate by parts twice and use the fact that the momenta are on-shell to obtain a boundary expression. Furthermore, when taking the large-$r$ limit of this expression we know from section~\ref{sec:softadditions}, that we can split each extrapolated operator into two terms
\begin{equation}
 \lim_{r\to\infty} r \, \phi=  \Phi_0  +  \Phi_{\textit{soft}},
\end{equation}
the two terms on the right hand side correspond to the first and second term of~\eqref{eq:corrected expression} respectively.\footnote{Furthermore, we can take the large-$r$ limit of the exponential without worrying about delta functions in $\omega$ as we have argued in section \ref{sec:softadditions} that the extrapolated operator does not give an IR divergence, so such a term would be proportional to $\omega \delta(\omega)$ and therefore vanish.}
To avoid clutter, we write down the resulting expression for a single operator, leaving it implicit that it is placed inside a correlation function. We get
\begin{equation}
  \lim_{r\to\infty} \int\limits_{-r\leq r' \leq r} d^4x \, e^{\pm i p\cdot x} \nabla^2 \phi =\frac{1}{2}\int du\,e^{\mp i \omega u }\, \left((\Phi^+_0 + \Phi^+_{\textit{soft}})-(\Phi^-_0 + \Phi^-_{\textit{soft}})\right),
    \label{eq:extrapLSZ}
\end{equation}
where the superscript denotes whether the operator is pushed to $\scri^+$ or $\scri^{-}$. It is important to keep track of these superscripts when we consider time ordering.

Let us first focus on contributions to the correlation function involving only non-soft operators. The Fourier transform picks out either the creation or annihilation operator depending on the sign in the exponent. Thus, when we impose time ordering, the only contribution that survives is the one where operators corresponding to the outgoing particle are extrapolated to the future and the ones corresponding to incoming particles are extrapolated to the past. The other contributions just annihilate the vacuum. In other words, the only non-zero contribution to the correlation function of this form is
\begin{equation}
\begin{split}
  \left(\int \prod_{i\in\textit{in}} du_i \, e^{-i \omega_i u_i}\right) &\left(  \int \prod_{j\in\textit{out}} du_j\,e^{+i \omega_j u_j}\right)\bra{0} \Phi_0^+(y_1)...\Phi_0^+(y_m)\Phi_0^-(x_1)...\Phi_0^-(x_n)\ket{0}\\
  &\propto\bra{0}a(p_1)...a(p_m)a^\dagger(p_1)...a^\dagger(p_n)\ket{0},
  \end{split}
\end{equation}
which is just the usual S-matrix element, as expected.  Let us then consider terms involving soft modes. Consider a correlation function that only contains a single soft factor. The soft factor is $u$-independent so the Fourier transform is proportional to a delta function in the frequency. Thus we would get schematically
\begin{equation}\label{eq:soft2}
   \delta(\omega) \left(\bra{0}\Phi^+_{\textit{soft}}\, S \ket{0}-\bra{0} S\, \Phi^-_{\textit{soft}}\ket{0}\right).
\end{equation}
 We first note that this expression has distributional support at $\omega=0$ and would not appear in the finite energy part of the S-matrix, or as a limit thereof.  It is straightforward to generalize to spinning massless external particles.  We thus see that the soft physics Ward identities remain in tact: these omitted soft contributions do not modify the Weinberg pole.

In gauge theories we also expect the full term in~\eqref{eq:soft2} to identically vanish. Each of the terms multiplying this delta function are proportional to the leading soft theorem~\cite{Weinberg:1965nx} (see~\cite{Kapec:2022axw,Derda:2024jvo} for recent discussions of soft scalars). Diagrammatically, one can see that the single soft insertion of an incoming vs outgoing particle differs only by a sign. Tracing the signs in~\eqref{eq:corrected expression} and \eqref{eq:soft2}, these contributions can be shown to cancel, so that~\eqref{eq:soft2} vanishes. Although it is less clear what happens for multiple soft insertions, one might hope that they cancel out in a similar way. We leave such an analysis for later work.

\subsection{Extracting Dynamics from Boundary Correlators}\label{sec:soft4pt}

In the previous section we reviewed how the extrapolate dictionary can be used to extract S-matrix elements, while also clarifying how the additional soft terms in our corrected extrapolate dictionary consistently drop out. We will now turn to the explicit example of $\lambda\phi^4$ theory to explore how these corrections can still capture interesting dynamics of the theory.  In particular, we will see that the correlation function of $\Phi_0$'s computes the amplitude, but that it does not match the naive extrapolated bulk correlation function, which has a time independent magnetic branch. Furthermore, we discuss how the magnetic branch captures the data of the low point correlators. 

First, let us start with the perturbative set up. Following a similar procedure to how we extracted the boundary-to-boundary propagator in section~\ref{sec:two_point} from the bulk-to-bulk propagator, we can get the bulk-to-boundary propagator by only sending one of the points to infinity. Let's start with just the $\Phi_0$ term from the previous subsection. We have
\be
\begin{split}
\langle T\{\phi(u,r,z,\bz) \Phi^\pm (u_0,z_0,\bz_0)\}\rangle =&\frac{1}{r|z-z_0|^2+u-u_0\pm i\epsilon} \\
=&\frac{1}{q(z_0,\bar{z_0})\cdot X(u,r,z,\bar{z}) -u_0\pm i\epsilon},
\end{split}
\ee
where $q(z_0,\bar{z}_0)=p(\omega,z_0,\bar{z}_0)/\omega$ and $X$ is the position of the bulk point. 
Now for this theory the three point correlator of single particle operators vanishes, so we will turn to the four point contact diagram. The connected correlator is then proportional to
\begin{equation}
\left\langle\Phi_0\left(x_1\right) \cdots \Phi_0\left(x_4\right)\right\rangle \propto \lambda I_0, \quad I_0 \equiv \frac{1}{2}\int d^4 X\,\prod_{i=1}^4 \frac{1}{\lambda_i q_i\cdot X -u_i\pm i\epsilon}.
\end{equation}
We can rewrite this integral using Schwinger parameters that replace 
\be
 \prod_{i=1}^4\frac{1}{q_i\cdot X -u_i\pm i\epsilon} =  \prod_{i=1}^4\int d\lambda_i e^{i\sum_i \lambda_i q_i \cdot X-\sum_i\lambda_i (u_i\mp i \epsilon)}.
\ee
Here the range of the $\lambda$ integral is $(0,\infty)$ or $(-\infty,0)$ depending on the $\pm i\epsilon$ prescription, i.e. whether the operator is extrapolated to future or past null infinity. If at this stage we Fourier transform from position $\{u_i\}$ to momentum space $\{\omega_i\}$ we see that we get factors of $\delta(\lambda_i\pm \omega_i)$ saturating the $\lambda_i$ integrals. The remaining integral over $X$ gives 
\begin{equation}\label{eq:phi4amp}
    I_0 \propto \delta^{(4)}\left(\sum_i \omega_i q_i\right),
\end{equation}
which is proportional to the momentum space amplitude. This is all consistent with our discussion of the non-soft part of the extrapolate dictionary above, and how the LSZ formulation lands on the S-matrix. However, we would like to restore the non-contact terms we would  expect from taking the flat limit of AdS/CFT. There is a long history of extracting S-matrix elements from the flat limit~\cite{Polchinski:1999ry,Penedones:2010ue,Gary:2009ae,Fitzpatrick:2010zm,Komatsu:2020sag}. We will focus on the bulk point configuration discussed in the recent works~\cite{Maldacena:2015iua,Hijano:2019qmi,Jain:2023fxc,Alday:2024yyj}. (See \cite{Hijano:2019qmi,deGioia:2022fcn,deGioia:2023cbd,deGioia:2024yne} for discussions that make direct contact with the celestial basis.

Before taking the large $r$ limit, one can evaluate the full four point function for four bulk points as in~\cite{Maldacena:2015iua}. For 
four spacelike separated scalars we have
\begin{equation}\label{fourbulk}
\left\langle\phi\left(x_1\right) \cdots \phi\left(x_4\right)\right\rangle \propto \lambda I, \quad I \equiv \int d^4 w \frac{1}{\prod_{i=1}^4\left[\left(w-x_i\right)^2+i \epsilon\right]}
\end{equation}
where
\begin{equation}
I=-\frac{2 \pi^2 i Z \bar{Z}}{x_{12}^2 x_{34}^2}\left[\frac{2 \operatorname{Li}_2(Z)-2 \operatorname{Li}_2(\bar{Z})+\log (Z \bar{Z}) \log \left(\frac{1-Z}{1-\bar{Z}}\right)}{Z-\bar{Z}}\right] 
\end{equation}
is a function of the cross ratios
\be
u=\frac{x_{12}^2 x_{34}^2}{x_{13}^2 x_{24}^2}=Z \bar{Z}, \quad v=\frac{x_{14}^2 x_{23}^2}{x_{13}^2 x_{24}^2}=(1-Z)(1-\bar{Z})
\ee
and the term in square brackets is the $D$-function
\begin{equation}
D_{1111}(Z, \bar{Z})=\frac{1}{Z-\bar{Z}}\left(2 \operatorname{Li}_2(Z)-2 \operatorname{Li}_2(\bar{Z})+\ln (Z \bar{Z}) \ln \left(\frac{1-Z}{1-\bar{Z}}\right)\right).
\end{equation}
These results are obtained by Wick rotating from Euclidean signature. Note that the function is not singular at $Z=\bar{Z}$. As one analytically continues to Lorentzian scattering kinematics the branch cuts in the logarithms will give rise to various phases that change the singularity structure. In particular, Wick rotating gives
\be\badat{3}
D_{1111}(Z, \bar{Z})&=\frac{1}{Z-\bar{Z}}\left(2 \operatorname{Li}_2(Z)-2 \operatorname{Li}_2(\bar{Z})+4 \pi i \ln \bar{Z} \right.\\
&~~~~~~~~~~~~~~~~~~~+\left.\left(\ln (Z \bar{Z})+2\pi i\right)\left( \ln \left(\frac{1-Z}{1-\bar{Z}}\right)-2\pi i\right)\right),
\eadat\ee
which now has a singularity when $Z=\bar{Z}$. This is analyzed in more detail in~\cite{Jain:2023fxc,Alday:2024yyj}.

The main point for our discussion is in regards to the behavior of the cross ratios $u,v$ appearing in the $D$ function as we go to large $r$. Namely writing
\be
x_i^\mu= u_i n^\mu + r q_i^\mu
\ee
where $n^\mu=(1,0,0,0)$ and $q_i(z_i,\bz_i)$ is a reference null vector we have
\be
(x_i-x_j)^2=r^2 |z_{ij}|^2+\mathcal{O}(r)
\ee
so that
\be
u\simeq \frac{|z_{12}|^2|z_{34}|^2}{|z_{13}|^2|z_{24}|^2},~~v\simeq \frac{|z_{14}|^2|z_{23}|^2}{|z_{13}|^2|z_{24}|^2}
\ee
are just the cross ratios on the celestial sphere. Taking the large $r$ limit of~\eqref{fourbulk} without being careful about contact terms lands us on a four point function that is analytic on the celestial sphere and $u_i$ independent. Now the stark contrast of this analytic part to the celestial amplitudes story was raise as an open question in~\cite{Jain:2023fxc}. Here we see that both terms naturally appear in the extrapolate dictionary and that the answer one gets by naively extrapolating the bulk four point function is a soft contribution that would usually be dropped.

We can actually have a little bit more fun.  Poincar\'e invariance guarantees that the scalar four point function should only depend on the $x_{ij}$.  Consider restricting these operators to the lightcone of the origin. The Lorentz group then acts as the 2D conformal group and the correlator should be decomposed into a function of the conformally invariant cross ratios and a kinematic factor determined by the 2D (Carrollian) scaling dimensions. These Carrollian weights are related to the bulk scaling dimensions, and will be consistent with the extrapolate dictionary. We would land on a magnetic branch Carrollian correlator. 

Now it is interesting to ask to what extent one can go the other way. Let us start with a magnetic branch correlator. If we naively promote every $z_{ij}\rightarrow r^{-2} x_{ij}$ of our magnetic branch correlator we would have a candidate bulk 4-point function that is Poincare invariant and should match the expected dimensional analysis with regards to powers of the radial coordinate. In the particular $\lambda \phi^4$ example above we would recover the bulk correlator. More amusingly we could then try to take the large $r$ limit keeping track of contact terms and should be able to get back the electric branch correlator corresponding to the finite energy part of the scattering amplitude. This relies on both analyticity of the position space correlator and us focusing on a low point scattering amplitude. The number of independent Mandelstam invariants for an $n$-point amplitude in 4D is $3n-10$. Meanwhile the number of independent cross ratios in a 2D CFT is $2n-6$. These are equal at $n=4$. However the four point function is all one needs to extract 2D CFT data, so this is not such a loss, especially if one can employ the aforementioned trick to recover the electric branch correlators. One should note that the soft magnetic branch correlators evade the constraints on the reality of the cross ratio that the electric branch 4 point function encounters (see~\cite{Mizera:2022sln} for a more detailed discussion of the kinematic constraints on celestial amplitudes at $n$-point).  

We conclude that, once we are equipped with some control over position space analyticity in the bulk, the magnetic branch correlators can encode a surprising amount of the dynamics of the bulk theory despite their trivial time dependence at the boundary!

\subsection{Connections to the Shadow Transform}\label{sec:shadow}

Let us close by connecting the above discussions about flat space extrapolate dictionaries to the literature on soft celestial amplitudes and shadow transforms. A good starting point is to look back at our expression~\eqref{eq:corrected expression} for the corrected extrapolate operator. We see that it takes the form of the naive extrapolate operator $\Phi_0$ plus a term that can be interpreted as a shadow transform of a $\Delta=1$ soft pole. Thus while we discussed in section~\ref{sec:LSZ} that the LSZ procedure can project out soft terms, one could try to formally try to extract the corrected correlators from finite energy S-matrix elements. 

Now consider the magnetic branch connected correlator we studied in section~\ref{sec:soft4pt}.  In the large $r$ limit the $u_i$-dependence drops out of the cross ratio, so $Z$ is purely a function of the coordinates $z_i$. As such, the naive extrapolated 4-point function will be proportional to $\prod_i \delta(\omega_i)$ when Fourier transformed in the $u_i$ coordinates. Similarly, the delta function \eqref{eq:phi4amp} has support on $\omega_i=0$, one might wonder if parts of the extrapolated correlator is captured by the soft support of the amplitude. It turns out that 
\begin{equation}
   \delta^{(4)}\left(\sum_i \omega_i q_i\right)\bigg|_{soft} \sim \frac{1}{|z_{12}|^2|z_{34}|^2(Z-\bar{Z})} \prod_i \delta(\omega_i),
\end{equation}
which is precisely the divergent part of extrapolated correlator. This is consistent with the fact, explained in \cite{Jain:2023fxc}, that the divergent term  in the extrapolated correlator will be proportional to the amplitude. It also is using the same observation that was noted in~\cite{Chang:2022jut,Chang:2022seh} when examining shadow celestial amplitudes in the soft limit. Namely: there are soft and colliner regions of the momentum conserving delta function besides those considered in investigations of hard amplitudes in~\cite{Pasterski:2016qvg,Pasterski:2017ylz}.

Finally, let us make some contact with the recent proposal regarding shadow celestial amplitudes in~\cite{Banerjee:2024yir}. In~\cite{Banerjee:2024yir} it was suggested that the analytic correlators in~\cite{Jain:2023fxc} were shadow celestial amplitudes. Mechanically that conclusion follows from dropping contact terms in the expansion of the conformal primary wavefunctions. As shown in~\cite{Donnay:2022sdg}, the series expansion of the shadow versus un-shadowed conformal primary wavefunctions exchange the leading contact and non-contact terms in a large $r$ expansion. These terms generally have different powers of $r$ except for special values of the celestial dimension. Curiously, at this conformal dimension a massive scalar field would become its own shadow. In the end our proposal for correcting the extrapolate dictionary involves shadow transforms in the soft limit, but in a different manner than the celestial proposal in~\cite{Banerjee:2024yir}. It would be interesting to explore a possible connection to the 3D shadow transforms used to extract soft Ward identities from the flat limit of AdS/CFT in~\cite{deGioia:2023cbd}.

\section*{Acknowledgements}

We are grateful to Luca Ciambelli, Finn Larsen, Justin Kulp, Prahar Mitra, Shiraz Minwalla, Richard Myers, Rob Myers, and Atul Sharma for useful discussions. The research of SP is supported by the Celestial Holography Initiative at the Perimeter Institute for Theoretical Physics and by the Simons Collaboration on Celestial Holography. SP and EJ's research at the Perimeter Institute is supported by the Government of Canada through the Department of Innovation, Science and Industry Canada and by the Province of Ontario through the Ministry of Colleges and Universities.

\appendix

\section{Details of the Regulated 2-Point Correlators}\label{app:extrapolatcheck}

In this section, we show that once we have taken care of the divergences in the extrapolate operator to arrive at \eqref{eq:extrapop}, we recover a boundary correlation function that is consistent with the extrapolated bulk correlation function. In particular, we recover the analytic part of the correlation function and replace the IR divergence in the boundary correlation function with a UV divergence that matches the bulk correlator. 

We split the extrapolated operator into $\Phi=\Phi_0+\Phi_{\textit{soft}}$ according to \eqref{eq:extrapop}. Let us first check the contribution to the 2-point function coming from the non-soft term $\langle\Phi_0\Phi_0\rangle$. The relevant momentum space commutation relation is 
\begin{equation}
    \begin{split}
        \big[\partial_{\omega_1} (\omega_1 &a(\omega_1,z_1,\bar{z}_1))-i \omega_1 u_1\,a(\omega_1,z_1,\bar{z}_1),\partial_{\omega_2} (\omega_2 a^\dagger (\omega_2,z_2,\bar{z}_2))+i \omega_2 u_2\,a^\dagger(\omega_2,z_2,\bar{z}_2)\big]=\\
        &
       \frac{1}{2}(2\pi)^{3} \left(-\omega_1 \partial_{\omega_1}^2-(1-i(\omega_1 u_1 + \omega_2 u_2))\partial_{\omega_1}+\omega_1 u_1 u_2\right)\delta(\omega_1-\omega_2)\delta^2(z_1-z_2),
    \end{split}
\end{equation}
which, using eq.~\eqref{eq:extrapop}, straightforwardly yields 
\begin{equation}\scalemath{.95}{
    \bra{0}\Phi_0(u_1,z_1,\bar{z}_1)\Phi_0(u_2,z_2,\bar{z}_2)\ket{0}= \frac{-1}{4\pi }\left(\gamma + \log\left(\omega_0 |u_1-u_2|)\right)+i \frac{\pi}{2}\text{sign}(u_1-u_2)\right)\delta^2(z_1-z_2). }
\end{equation}
This result looks similar to~\eqref{eq:IRdivversion}, with one difference: the factor $\omega_0$ was an arbitrary dimensionful constant we introduced when cancelling the IR divergence by integration by parts, rather than an IR cutoff that we used to regulate the divergence. If we choose $\omega_0$ to match the contact term in the position space expression \eqref{eq:UVdivversion}, namely $\omega_0=e^{-\gamma}r/\epsilon$, we would find it to be a UV scale. Interestingly, this is different from the previously obtained expression \eqref{eq:IRdivversion}, which contains an IR divergence. 

To compute the cross term contribution coming from $\langle\Phi_0 \Phi_{\textit{soft}}\rangle$ we need the momentum space commutation relation 
\begin{equation}
\begin{split}
    \left[\partial_{\omega_1} (\omega_1 a(\omega_1,z_1,\bar{z}_1))-i \omega_1 u_1\, a(\omega_1,z_1,\bar{z}_1),\,\omega_2 \partial_{|z_2|}\, a^\dagger (\omega_2,z_2,\bar z_2) \right]=\\
    -\frac{1}{2}(2\pi)^3 \left(1-i \omega_1 u_1+\omega_1 \partial_{\omega_1}\right)\delta(\omega_1-\omega_2)\,\partial_{|z_1|}\delta^2(z_1-z_2),
\end{split}
\end{equation}
which yields 
\begin{equation}
    \bra{0}\Phi_0(u_1,z_1,\bar{z}_1)\Phi_{\textit{soft}}(u_2,z_2,\bar{z}_2)\ket{0}= \frac{1}{8\pi^2}\frac{1}{|z_1-z_2|^2}
\end{equation}
giving us the analytical part of the correlation function. Finally, the contribution from $\langle\Phi_{\textit{soft}}\Phi_{\textit{soft}}\rangle$ is seen to vanish, as we have the corresponding commutation relation
\begin{equation}
     \left[ \omega_1 \partial_{|z_1|}\, a (\omega_1,z_1,\bar z_1),\,\omega_2 \partial_{|z_2|}\, a^\dagger (\omega_2,z_2,\bar z_2) \right]= \frac{1}{2}(2\pi)^3 \omega_1 \delta(\omega_1-\omega_2) \,\partial_{|z_1|}\partial_{|z_2|}\delta^2(z_1-z_2),
\end{equation}
which contains a factor $(\omega_1-\omega_2)\delta(\omega_1-\omega_2)$. 
Putting all of the above together, we find 
\begin{equation}\begin{split}
    \langle \Phi(u_1,z_1,\bar{z}_1)\Phi(u_2,z_2,\bar{z}_2)\rangle &= \frac{1}{4\pi^2}\frac{1}{|z_1-z_2|^2}-\frac{1}{4\pi}\left(\gamma + \log\left(\omega_0 (u_1-u_2)\right)+i \frac{\pi}{2}\right)\delta^2(z_1-z_2). 
    \label{eq:newcorr}
\end{split}\end{equation}
We have thus obtained both the analytic and the distributional parts of the 2-point function.

\section{Extrapolating in Different Directions}\label{app:extrapolate}

When extrapolating correlation functions from the bulk to the boundary, we should foliate spacetime by a one parameter family of codimension one surfaces such that when we take the parameter to infinity while keeping the other coordinates fixed, we push the operators to null infinity. Here we note one apparent difference between whether one extrapolates on null, timelike or spacelike slices. Namely, whether there is a $u$-dependent delta function appearing in the two-point function or not when we push both operator to future (or past) null infinity.

\paragraph{Timelike slicing:}
In the case of extrapolating on timelike surfaces, we choose to work with spherical Bondi coordinates. 
$(\tilde{u},\tilde{r},\tilde{z},\bar{\tilde{z}})$ given by
\begin{equation}
    x^{\mu}=\left(\tilde{u}+\tilde{r},\frac{\tilde{r}(\tilde{z}+\bar{\tilde{z}})}{1+|\tilde{z}|^2},-\frac{i\tilde{r}(\tilde{z}-\bar{\tilde{z}})}{1+|\tilde{z}|^2},\frac{\tilde{r}   (1-|\tilde{z}|^2)}{1+|\tilde{z}|^2}\right),
    \label{eq:Spherical bondi}
\end{equation}
for which the metric becomes 
\begin{equation}
        ds^2= -d\tilde{u}^2-2d\tilde{u}\,d\tilde{r} +\frac{4\tilde{r}^2}{(1+|\tilde{z}|^2)^2}d\tilde{z}\,d\bar{\tilde{z}}.
\end{equation}
This foliates the spacetime into timelike surfaces of constant $r$-coordinate, onto which we will place all operators inside a correlation function before taking the extrapolate $r\to\infty$ limit. Note that here we are showing the retarded Bondi coordinates which are adapted to extrapolation to $\scri^+$, while the advanced coordinates obtained by replacing $\tilde{u}+\tilde{r} \to \tilde{v}-\tilde{r}$ are adapted to $\scri^-$.

\paragraph{Spacelike slicing: }
For spacelike extrapolation a convenient coordinate system is obtained from eq.~\eqref{eq:Spherical bondi} by changing coordinates to $\tilde{r}\to\tilde{t}-\tilde{u}$ such that the metric becomes
\begin{equation}\label{eq:spacelikeslice}
    ds^2=-2d\tilde{t}d\tilde{u}+d\tilde{u}^2+\frac{4(\tilde{t}-\tilde{u})^2}{(1+|z|^2)^2}d\tilde{z}d\bar{\tilde{z}},
\end{equation}
which gives a foliation of spacetime by surfaces of constant $\tilde{t}$-coordinate. In this case, one performs the extrapolation by putting all the operators in a correlation function on such a surface before taking $\tilde{t}\to\infty$. As in the timelike case, the above are adapted to extrapolation to $\scri^+$, while a change of coordinates $\tilde{t}-\tilde{u}\to\tilde{t}+\tilde{v}$ is adapted to $\scri^-.$

\paragraph{Null slicing:}
For null extrapolation a natural choice is Bondi coordinates $(u,r,z,\bar{z})$, as we used above in \eqref{eq:flat-bondi}. To reiterate, these  coordinates are given by
\begin{equation}
    x^{\mu}=\frac{r}{2}\left(1+|z|^2+\frac{u}{r},z+\bar{z},-i(z-\bar{z}),1-|z|^2-\frac{u}{r}\right), \label{eq:flatbondi}
\end{equation}
such that the metric becomes
\begin{equation}
    ds^2=-du\,dr+r^2\,dz\,d\bar{z}.
\end{equation}
In these coordinates future null infinity lies at $r\to\infty$ and past null infinity lies at $r\to-\infty$ and the surfaces with constant $r$-coordinate are null.

\begin{figure}[ht]
\centering
\vspace{-0.5em}
\begin{tikzpicture}[scale=2.1]
\definecolor{darkgreen}{rgb}{.0, 0.5, .1};
\draw[thick, red] (0,1) to [bend left=45] (0,-1);
\draw[thick, blue] (0,.5) to [bend left=22] (1,0);
\draw[thick,dashed, blue] (0,-.5) to [bend right=22] (1,0);
\draw[ultra thick, gray] (0,.9) -- (1-.05,-0.1+.05);
\draw[thick, dashed, gray] (0,-.9) -- (1-.05,0.1-.05);
\draw[thick] (0,-1) -- (1,0) node[right]{$i^0$} -- (0,1) ;
\draw[thick] (0,-1) node[below]{$i^-$}  -- (0,1) node[above]{$i^+$};
\end{tikzpicture}
\caption{Extrapolating to the conformal boundary with operators on timelike (red), spacelike (blue), and null (gray) hypersurfaces in $\mathbb{R}^{1,3}$. Outgoing and incoming are represented by solid and dashed lines respectively. For the timelike surface only the window where $u=t-r=\mathcal{O}(R^0)$ maps to $\mathcal{I}^+$, and similarly for $v=t+r=\mathcal{O}(R^0)$ and $\mathcal{I}^-$.
}
\label{fig:inversion}
\end{figure}
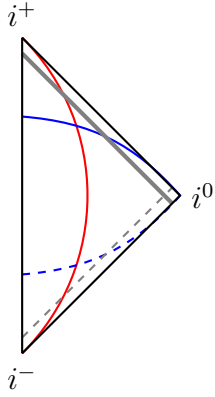

\vspace{1em}

 In section~\ref{sec:carrollbranches} we examined the case of operators placed on null hyperplanes near future and past null infinity. We avoided the case of placing both operators on the same light sheet. This would be relevant for computing vacuum wavefunctionals rather than S-matrix elements. Let's now discuss what happens. When both operators are on the same null hyperplane, it is unclear how to regulate the contact term. 
Namely,
\begin{equation}
    \begin{split}
        \langle T\{r_1\phi(x_1) r_2\phi(x_2)\} \rangle|_{r_1=r_2} &= \frac{1}{4\pi^2}\frac{r^2}{(x_1-x_2)^2+i\epsilon}\\
        &= \frac{1}{4\pi^2}\frac{1}{|z_1-z_2|^2+i\epsilon} ,
    \end{split}
    \label{eq:++corr}
\end{equation}
where we have absorbed a factor of $r^2$ in $\epsilon$. In particular there is no $u$ dependence in the denominator. Either the timelike or spacelike regulator let us use the $\tilde{u}$ separation to regulate $(x-y)^2$ when the particles are collinear, and thus evaluate the contact term. Compare this to a timelike regulating surface, which gives 
\begin{equation}
    \begin{split}
        \langle T\{\tilde{r}_1\phi(x_1) \tilde{r}_2\phi(x_2)\} \rangle|_{\tilde{r}_1=\tilde{r}_2=\tilde{r}} &= \frac{1}{4\pi^2}\frac{\tilde{r}^2}{(x_1-x_2)^2+i\epsilon}\\
        &= \frac{1}{4\pi^2}\frac{1}{\frac{4|z_1-z_2|^2}{(1+|z_1|^2)(1+|z_2|^2)}-\left(\frac{u_1-u_2}{\tilde{r}}\right)^2+i{\epsilon}}.
    \end{split}
    \label{eq:++corr}
\end{equation}
For $z_1\neq z_2$ the large $r$ limit is straightforward to obtain. When $z_1=z_2$ we see that the large $r$ limit diverges in a $u$ dependent way. Indeed, we can discover the presence of a $u$-dependent delta function term by considering the difference between \eqref{eq:++corr} and the naive large $\tilde{r}$ limit and integrating over $z$. The leading contribution in $\tilde{r}$ is
\begin{equation}
    \begin{split}
   \, c(u)\equiv&\,\frac{1}{8}\left(i \pi + 2 \arctan\left(\frac{\epsilon}{u^2}\right)- \ln\left(\frac{u^4+\epsilon^2}{\epsilon^2}\right)\right)\\
   &= \begin{cases}
   \frac{i \pi}{4}\qquad\qquad\qquad\qquad\quad\, \text{if $u=0$,}\\
   \frac{1}{2}\left( \frac{i\pi }{4 }-\ln\frac{|u|}{\sqrt{\epsilon}}\right)+O(\epsilon)\quad \text{otherwise}.
   \end{cases}
    \end{split}
\end{equation}
Interpreting this as the presence of a delta function at $z_1=z_2$, we find that the expression for the extrapolated correlation function is 
\begin{equation}
\lim_{\tilde{r}\to\infty}\langle T\{\tilde{r}_1\phi(x_1) \tilde{r}_2\phi(x_2)\} \rangle|_{\tilde{r}_1=\tilde{r}_2=\tilde{r}} =\frac{1}{16\pi^2
} \frac{(1+|z_1|^2)(1+|z_2|^2)  }{|z_1-z_2|^2+i\epsilon}+\frac{1}{4\pi
} c(u_1-u_2) \delta^{(2)}(z_1-z_2).
\label{eq:timelikeextr}
\end{equation}
This is the expected CFT form of the two point function (on a sphere) plus a delta-function term. Extrapolation on a spacelike slices ends up looking identical to the timelike case above. In conclusion, we see that when we extrapolate both operators to the future/past, the choice of extrapolation prescription appears to affect whether there is a $u$-dependent contact term or not.


\bibliographystyle{jhep}
\bibliography{bibliography}

\end{document}